# Balancing Creativity and Automation: The Influence of AI on Modern Film Production

XU YIREN[1]

[1]Faculty of Science, Universiti Malaya, Kuala Lumpur, Malaysia

## Abstract

The integration of artificial intelligence (AI) into film production has revolutionized efficiency and creativity, yet it simultaneously raises critical ethical and practical challenges. This study explores the dual impact of AI on modern cinema through three objectives: defining the optimal human-AI relationship, balancing creativity with automation, and developing ethical guidelines. By employing a mixed-method approach combining theoretical frameworks (auteur theory, human-technology relations) and case studies (The Safe Zone, Fast & Furious 7, The Brutalist), the research reveals that positioning AI as an "embodiment tool" rather than an independent "alterity partner" preserves human authorship and artistic integrity. Key findings highlight the risks of surveillance capitalism in AI-driven markets and the ethical dilemmas of deepfake technology. The study concludes with actionable recommendations, including international regulatory frameworks and a Human Control Index (HCI) to quantify AI involvement. These insights aim to guide filmmakers, policymakers, and scholars in navigating the evolving AI-cinema landscape while safeguarding cultural diversity and ethical standards.

## 1. Introduction

In recent years, the rapid development of artificial intelligence (AI) has revolutionized the film industry, with its applications ranging from subtle to overt across various stages of production and dissemination. AI has not only enhanced the efficiency of film production but also brought transformative changes to how films are distributed and consumed (Zhao & Zhao, 2024). By integrating human creativity with AI-driven tools, filmmakers have experienced significant advancements in every phase of production—from scriptwriting and pre-production planning to post-production editing (Peiming, 2023; Fang, 2023). For instance, AI-powered software can assist screenwriters in generating story ideas, optimizing shooting schedules, and even enhancing visual effects (Izani et al., 2025).

In addition to its role in production, AI has reshaped the landscape of film dissemination. Initially, AI was often seen as a gimmick, used primarily to attract audiences' interest through novelty (Miller, 2024). However, this perception has evolved significantly. Traditional methods of advertising films, which required substantial time and labor costs, have been replaced by AI-driven marketing tools that enable filmmakers to target specific audiences with precision. These tools not only reduce costs but also maximize promotional impact (Zhao & Zhao, 2024; Lu et al., 2023). Furthermore, AI algorithms can analyze audience preferences and predict box office performance, providing valuable insights for studio decision-making (Akbari & Jastacia, 2024).

Despite these advancements, the integration of AI into the film industry has raised several ethical and practical challenges. One key issue is the question of authorship and ownership in films created through human-AI collaboration (Zhu & Zhang, 2022). For example, who should be credited as the creator of a scene or script generated by an AI tool? Additionally, the misuse of deepfake technology poses significant risks to both the film industry and society at large (Bustillos et al., 2023). These challenges highlight the need for clear guidelines and regulations to govern the use of AI in filmmaking.

This study aims to address these issues by exploring three main objectives:

1. Indicate the suitable the suitable relationship between AI and human.
   What's the suitable role of AI in film production? A tool or an independent "alterity partner"?
2. Balancing human creativity and AI automation in film production.
   How can filmmakers integrate AI tools without compromising artistic integrity or human agency?
3. Developing ethical guidelines for AI use in the film industry.
   How can stakeholders mitigate risks such as deepfake misuse while maximizing the benefits of AI?

By addressing these questions, this research seeks to provide actionable recommendations for filmmakers, studios, and policymakers to navigate the rapidly evolving landscape of AI in cinema.

## 2. Literature Review

### 2.1. AI in Film Production

There has been significant research on the application of AI in film production. The majority of studies have focused on how AI technology can empower film production, particularly in the visual domain. For instance, researchers have explored how AI can generate high-quality images and videos (Song et al., 2024). Additionally, several studies have examined the collaboration between AI and film production, such as how AI can assist various stages of filmmaking, from idea generation to enhancing visual effects (Zhao & Zhao, 2024; Peiming, 2023; Fang, 2023).

The advantages of AI in film production are evident. Its ability to reduce labor and time costs makes it a promising solution to the challenges traditionally faced by the film industry. However, as Xiao Gong highlights in his research, there is a growing need for interdisciplinary talent to fully harness the potential of AI in this field (Gong, 2024). Despite some limitations in the maturity of certain AI technologies, the integration of AI into film production is undeniably an essential direction for the future (Yu et al., 2024).

Moreover, while AI has already made inconspicuous contributions to film production, there have been instances where the use of AI-generated or AI-assisted elements in films was downplayed (Chow, 2020). Nevertheless, recent examples demonstrate that AI's role in filmmaking is no longer just a theoretical concept. For instance, The Safe Zone utilized ChatGPT for scriptwriting, camera movement instructions, lighting requirements, and wardrobe suggestions, with storyboards generated by DALL-E (as cited in "AI list: The best (and weirdest) AI-generated films," 2023). This film was completed just 17 days after the release of ChatGPT. Similarly, director Dávid Jancsó acknowledged the use of AI to refine Hungarian dialogue for his film The Brutalist (T- Labs Team, 2025).

Furthermore, AI deepfakes have become increasingly prevalent in substituting makeup and representing actors of different ages or body shapes, as well as reviving deceased actors. Notable examples include The Irishman, Central Intelligence, and Fast & Furious 7 (teresa_myers, 2024). Despite these advancements, research on how to effectively integrate AI automation with human creativity remains underdeveloped. This gap is particularly significant given the current era of AI evaluation in the film industry. Therefore, this study aims to provide practical guidelines for balancing the collaboration between AI and human creativity in film production.

## 2.2. AI-generated or AI-assisted Films in Market

On the one hand, scholars have demonstrated that AI can be utilized to meet audiences' demands for interactivity, immersion, and virtuality in media-driven art practices,which is driven by data- and algorithm-powered content distribution revolutions and the rise of AI-generated content (AIGC) (Fang, 2023; Liu & Li, 2024). On the other hand, AI can enhance its influence in real-world markets through algorithm-driven content distribution and precise promotion on social media platforms (Ozgun & Treske, 2024; Tyagi, 2020).

Additionally, according to Zhang and Shao's PSO-BP model, emerging technologies such as 5G can be integrated with AI to elevate film production and dissemination to a higher level (Zhang & Shao, 2025). While these scholars acknowledge that there are some experimental errors in their results, the discrepancy between the estimated and actual values is minimal. This suggests that their overall assumptions are valid, even if certain small parameters in their model require refinement.

Moreover, while integrating AI into film production can yield numerous benefits in real-world markets—such as predicting a film's commercial viability and optimizing marketing strategies (Swarnakar, n.d.)—there is currently no existing literature specifically addressing the actual market performance of AI-generated or AI-assisted films. The only available data come from isolated examples. For instance, Fast & Furious 7, an AI-assisted film that utilized AI deepfakes, grossed $1,515,047,671 worldwide during its original release ("Furious 7," n.d.). In contrast, The Safe Zone, the first film directed and written entirely by AI, garnered 123K views, 692 likes, and 76 comments on Facebook ("Safe Zone: First Film DIRECTED & WRITTEN by Artificial Intelligence (AI)," 2022). To better evaluate the dissemination effects of AI-generated and AI-assisted films in real-world markets, this research will select representative cases, gather specific data, and conduct qualitative analyses of online reviews.

## 2.3. Ethical Issues

Whether or at which level AI can generate creativity remains one of the most fundamental ethical issues raised by the use of AI in films. While recent studies have demonstrated that current AI systems can collaborate with human creativity, they are still insufficient to produce original creativity independently, as their outputs are limited to what is derived from their training data (Zhu & Zhang, 2022). In academia, scholars hold divergent views on this matter: some argue that AI can enhance creative liberation, while others contend that it poses a threat to artistic uniqueness (Garcia, 2024; Kirkpatrick, 2023). This debate remains highly contentious.

Beyond creativity, the question of whether AI challenges authorship and ownership in film production is equally significant. For instance, scripts assisted by AI—whether through polishing or comprehensive generation—blur the boundaries of authorship and ownership, as the final work reflects both the values and desires of the human creator and the AI (Islam & Greenwood, 2024; Weber et al., 2024). Furthermore, since current AI systems can only generate content based on their training data, scholars have raised concerns that this could lead to a homogenization of creative output, potentially undermining originality and diversity (Zhu & Zhang, 2022).

In terms of responsibility, the issue becomes even more complex. For instance, deepfakes may infringe individuals' rights, particularly portrait rights (Visser, 2025). While numerous laws relevant to AI have been established—such as the Copyright in the Digital Single Market Directive (CDSMD) in the European Union, which outlines responsibility divisions for AI-related issues—these legal frameworks still face challenges when applied to real-world scenarios ("Directive - 2019/790 - EN - DSM - EUR-Lex," n.d.). Specifically, deepfakes created under the category of neighboring rights remain inadequately addressed in existing laws (Visser, 2025).

From an ethical standpoint, according to moral subjectivity, moral responsibility lies with the users or developers (orderers) rather than AI itself for its actions (Wróblewski & Fortuna, 2023). This is because current AI lacks free will and cannot operate independently without external control from its users. However, as future AI systems evolve to possess the ability to learn, adapt, and make decisions autonomously (Decker, 2017), there may come a time when AI could be considered a morally responsible entity and held accountable for its actions.

Finally, the question of bias cannot be overlooked. AI systems are not immune to the biases inherent in their training data, which could result in unintended discriminatory outcomes in film production (Weber et al., 2024). Addressing these challenges requires a multidisciplinary approach that involves ethicists, legal experts, and technologists working together to ensure that AI is used responsibly and ethically in the film industry.

## 3. Theoretical Framework

### 3.1. Insuring Human Auteurship: The Role of AI in Film Production

The integration of AI into film production poses a significant challenge to the foundational principles of Sarris' auteur theory. According to this theory, directors, as the "authors" of films, should exhibit signature styles and construct personal worldviews through their works (Sarris, 1963). However, the rise of AI-generated and AI-assisted films undermines this framework, as directors may be heavily influenced by AI systems during production. This is because AI synthesizes existing visual languages to create new ones, often blurring the distinction between human creativity and machine-generated output (Reddy et al., 2024).

For instance, The Safe Zone utilized AI in various aspects of its production, including scriptwriting, camera movement, lighting, wardrobe, and storyboarding ("AI list: The best (and weirdest) AI-generated films," 2023). It was even marketed as the "First Film DIRECTED & WRITTEN by Artificial Intelligence" ("Safe Zone: First Film DIRECTED & WRITTEN by Artificial Intelligence (AI)," 2022). However, since AI generates content based solely on its training data—derived from human works—the resulting film lacks a distinct directorial style and artistic coherence. This raises concerns about the erosion of authorship in an era where AI increasingly influences creative processes.

To address this issue, it is essential to ensure that the director remains a human entity when incorporating AI into film production. In Sarris' auteur theory, signature styles encompass visual aesthetics, recurring themes, and consistent use of cast members (Sarris, 1963). To preserve authorship while leveraging AI, directors should train their AI systems using their previous works. By doing so, AI can learn and adapt to the director's unique style, enabling it to serve as a tool that enhances rather than replaces human creativity. This approach ensures that the final product reflects the director's artistic vision while maintaining the integrity of authorship in an increasingly AI-driven industry.

### 3.2. Alter Partner or Embodiment Tool: The Relationship between Human and AI

The relationship between humans and AI in film production can be analyzed through Ihde's framework of Human-Technology Relations. Specifically, both the "embodiment relation" and "alterity relation" can describe this dynamic under different circumstances, but a critical question arises: which relationship is most appropriate? For instance, in Fast & Furious 7, the relationship is alterity relation, as the audience know Paulr Walker has died, while in The

Brutalist, the relationship is embodiment relation, as the audience couldn't know AI has refined its Hungarian dialogue if the director haven't stated (T- Labs Team, 2025; eresa_myers, 2024; Ihde, 1991).

To safeguard the ontological integrity of cinema, this essay argues that AI should be positioned as an "embodiment" tool rather than an independent "alterity partner". By framing AI as a tool that extends human creativity rather than as an autonomous entity, filmmakers can maintain human agency in the creative process (Ihde, 1991). This approach not only preserves the essence of cinematic artistry but also avoids potential ethical dilemmas. For example, moral accountability would then rest with the specific individual employing AI in their work, ensuring clarity and responsibility in the creation process.

Compared with embodiment relation, alterity relation, which regard AI as an independent "alterity partner", ignore the artistry of cinema. Though there are four theories on the origin of art, the main point is that art was invented by human, and human artists are the origin of artificial works (Heidegger, 2017). By positioning AI as an extension of human creativity rather than an independent force, filmmakers can navigate the complexities of technological integration while preserving the core values of artistic expression.

### 3.3. Protecting Original Film Market: The Surveillance Capitalist Threat of AI in the Film Market

Zuboff's theory of surveillance capitalism reveals a fundamental tension in AI-driven film markets: the extraction of "behavioral surplus" from audiences and creators to fuel predictive commodification (Zuboff, 2023). When studios deploy AI to generate scripts based on social media trends or optimize promotions through algorithmic targeting, cinema is reduced to a "prediction market"—where artistic value is subordinated to datafied consumer preferences (Zuboff, 2023). This logic disproportionately harms independent films, which lack the resources to compete in an ecosystem privileging algorithmic scalability over niche originality. Besides, as developed countries and economic powers own more powerful AI tools, this will absolutely break the balance between states and markets, and this is also an essential reason in the rise of surveillance capitalism (Haggart, 2019).

In this way, AI is not just a tool help us to forecast box office, but a weapon of surveillance capitalism, which will exploit both the directors and the audiences. In the one hand, as the script is generated based on the public's interest to gain more box office, the directors will be forced to cater to the market, and can't express their unique thought of the world (Mirowski et al., 2023). On the other hand, as the public gets used to purchasing by algorithm recommendation, they will be controlled by the owner of AI algorithm, and they will lose their right of privacy (Tucker et al., 2018). Additionally, the independent films will lose their only market.

As a result, we should protect original film market by pushing surveillance to against the threaten taken by the abuse of AI. This section provides four suggestions as bellow to help with the surveillance:

1. Establishing international regulatory bodies: These organizations would oversee AI-driven tools in the film industry, ensuring transparency and fairness in their application.

2. Developing legal frameworks: Laws should be enacted to regulate algorithmic recommendations, ensuring that they do not disproportionately favor mainstream content over independent works.

3. Promoting non-profit oversight organizations: Independent watchdogs could monitor AI's impact on cultural markets and advocate for policies that protect artistic diversity.

By addressing the challenges posed by surveillance capitalism in the film industry, we can create a more equitable ecosystem that supports both commercial and independent filmmakers.

## 4. Case Study Analysis

### 4.1. Case Selection Criteria

To systematically evaluate the dissemination effects of AI-generated and AI-assisted films, this study adopts a stratified case selection framework based on three dimensions: technological integration, market performance, and cultural legitimacy. The selected cases must meet the following criteria:

1. AI Intervention Scope: the film must utilize AI in at least two core production stages (e.g., scriptwriting, visual effects, editing) or one stage with significant artistic impact (e.g., deepfake-driven performance);

2. Data Accessibility: publicly available box office records and social media engagement metrics (views);

3. Representative Significance: includes both mainstream and independent films to reflect AI's differential impacts across budget scales and cultural contexts.

Based on these criteria, three cases were selected:

**Table 1**

*Cases Selection Criteria*

| Film Name | AI Application | Representative Value |
|---|---|---|
| The Safe Zone | Full AI Workflow | First Film Directed and Written by AI |
| Fast & Furious 7 | Deepfake Paul Walker | Benchmark for AI's Commercial |

|                | The Brutalist | Hungarian Dialogue Refined by AI | Remediation AI-assisted Independent Film |
|---|---|---|---|

*Note. References:*

*Teresa_Myers. (2024, July 29). Deepfake in movies | Facts, fiction & future of film industry. Facia.ai. https://facia.ai/blog/deepfake-into-movies-facts-fiction-future-of-film-industry/*

*T-Labs Team. (2025, January 27). How will AI change filmmaking? These 6 movies show how the technology has been used. Tatler Asia. https://www.tatlerasia.com/lifestyle/entertainment/ai-in-movies*

*The AI list: The best (and weirdest) AI generated films. (2023, December 24). SPYSCAPE | NYC's #1 rated museum & experience. https://spyscape.com/article/ai-film-roundup*

This stratified approach ensures methodological rigor while capturing AI's multifaceted role in contemporary cinema.

## 4.2. AI in Production: Comparative Analysis of The Safe Zone, Fast & Furious 7 and The Brutalist

There three cases exemplify different modes of AI's participation in film production, reflecting tensions between creative autonomy and technological determinism.

The Safe Zone: Full Automation as Artistic Paradox

As the fist film written and directed by AI, the production of The Safe Zone highlights the erosion of traditional auteurship. It's the lack of human director makes this film lose its artistry, which has been mentioned in Garcia's research (Garcia, 2024) Besides, while ChatGTP generates its script based on the previous scripts, the script of this film loses originality.

Fast & Furious 7: Deepfakes as Remediation Tool

The digital resurrection of Paul Walker after his death through deepfakes illustrates Ihde's "alterity relationship". The audience has known his death, and consciously engaged AI as an embodiment of him. Additionally, the studio strategically framed his deepfake as an "embodiment" of his legacy to improve audience reception. It's the multifunction of AI makes its box office grossed $1,515,047,671 worldwide during its original release ("Furious 7," n.d.). Besides, the success of its box office also revealing commercial cinema's instrumental logic on AI.

The Brutalist: Curated Collaboration

In contrast, The Brutalist only used AI for refining its Hungarian dialogue (T- Labs Team, 2025). In this case, the director input the film's original Hungarian dialogue, and asked AI to make it localized (T- Labs Team, 2025). As a result, the auteurship of this film is also its director, while AI is still a tool, but not another independent "alterity partner" (Ihde, 1991). This indicates us a way to balance human creativity and AI automation.

**Table 2**

*AI's Impacts on the Tree Cases*

| Film | AI's Role | Relationship between AI and Human | Auteurship of the Film | Box Office(USD) or Viewers(V) |
|---|---|---|---|---|
| The Safe Zone | Writer & Director | Alterity Relation | AI | 242K V |
| Fast & Furious | Post-production Tool | Alterity Relation | Human | $1,515,047,671 |
| The Brutalist | Localization Assistant | Embodiment Relation | Human | $41,461,248 |

Note. Reference:

The brutalist. (n.d.). Box Office Mojo. https://www.boxofficemojo.com/title/tt8999762/

Furious 7. (n.d.). Box Office Mojo. https://www.boxofficemojo.com/title/tt2820852/

[@Richard Juan]. Safe Zone: First Film DIRECTED & WRITTEN by Artificial Intelligence (AI) [FIRST FILM DIRECTED AND WRITTEN BY ARTIFICIAL INTELLIGENCE]. (2022, December 17). Facebook. https://www.facebook.com/richardjuan/videos/the-safe-zone-first-film-directed-written-by-artificial-intelligence-ai/1106246483382930/

## 5. Discussion& Conclusion

### 5.1. Balancing Creativity and Automation

Reviewing these three cases we selected, the role of AI could be both embodiment tool and an independent "alterity partner". However, regarding AI as one of the film artists is against the origin of art (Heidegger, 2017). Films which use AI as an alterity partner belongs to generative art, but not film itself (Boden & Edmonds, 2009). In order to insure human auteurship and the ontology of cinema, AI should be used as an embodiment tool, but be considered as an independent "alterity partner".

Besides, as presently we don't know the level of AI used in film production, we need to quantify it. For instance, the director of The Brutalist announced he used AI to refine its hungarian dialogue (T- Labs Team, 2025), but we still couldn't know at which level he used it. Just some difficult part or the whole? Though maybe we could ask individuals to announce more clearly in the future, we will still don't know if they lie. As a result, the surveillance is of great importance, and we need to quantify it.

In addition to the need of developing relevant algorithms to identify AI autonomy, this section rises a theory to help with the quantification during the surveillance:

**Figure 1**

*Human Control Index (HCI) to Evaluate the level of AI Use*

$$\text{Human Control Index (HCI)} = \frac{\text{Director's Decision Power}}{\text{AI Autonomy}}$$

*Note. Figure by Author*

When the result approaches zero, the film is completely made by AI; when the result approaches infinity, the film is completely made by human. However, as the lack of data, this research just rises a way to evaluate the level of AI use in quantification.

To summarize, this section offers five recommendations to help balance creativity and automation in film production:

    1. Ensure the Role of AI: Recognize that AI serves as a supportive tool rather than an autonomous entity or partner in the creative process. It is essential to maintain human oversight and decision-making to preserve artistic integrity;

    2. Clarify Individual Transparency: Encourage individuals and teams to provide transparent reports on the extent of AI utilization in their work. Clear communication about AI involvement helps ensure accountability and fosters trust among stakeholders;

    3. Establish a Regulatory Authority: Create an independent authority responsible for monitoring and regulating AI usage in film production. This entity should enforce ethical guidelines, assess compliance, and address potential misuse or over-reliance on automation;

    4. Develop Advanced Algorithms: Invest in the creation of algorithms designed to identify and manage AI autonomy effectively. These tools should be capable of detecting unauthorized or excessive AI involvement in creative processes while maintaining a balance between efficiency and human creativity;

    5. Quantify AI Usage with HCI Frameworks: Implement Human-Computer Interaction (HCI) frameworks to quantify and monitor AI involvement throughout the production process. This approach ensures that automation is used judiciously, aligning with artistic goals and preserving the human element in filmmaking.

By implementing these recommendations, the film industry can strike a harmonious balance between leveraging AI for innovation and maintaining the essential role of human creativity and oversight.

### 5.2. Ethical Guidelines for AI in Film Production

Though the deepfake of Paul Walker was advertised as the "embodiment" of his legacy by the studio of Fast & Furious 7, there were also plenty of audiences disagree with this ("Furious 7 used wayyyy more face replacement CG than I thought. They used it for 260 shots," 2019). The statement of "legacy" is very hypocritical, maybe Paul also disagree with it if he was alive. This is the most aggressive ethical problem in this film——using the dead as stunt. As a result, we need the permission from people alive, if there's a need for deepfake, while relevant laws should be enhanced.

Besides, compared with the box office of Fast & Furious 7 and The Brutalist, there's also a huge gap between commercial film and independent film. One of the most importance reasons is the accurate diversion led by AI algorithm bias during the promotional phase of the film. This bias has broadened the gap between the mainstream content and the niche content, and will absolutely make film market into a "prediction market", which will make independent film suffer more (Zuboff, 2023). Additionally, this will also manipulate the audience's consumption of movies and even their thinking. In order to cope with this, we need not only to reduce AI algorithm bias, but also to enhance our education systems to cultivate the ability of the public to think critically.

This section presents four recommendations aimed at addressing the challenges posed by deepfakes and other AI technologies in filmmaking:

1. Improve Relevant Laws: It is essential to establish clear and specific laws governing deepfakes and other AI technologies. These laws should define precise boundaries and metrics to ensure accountability and ethical use;

2. Obtain Personal Permission for Deepfakes: Before using deepfakes, it is crucial to obtain personal consent from individuals while they are alive. Additionally, there should be no false claims of deepfakes being used as a "legacy." Ensuring transparency in the creation and use of deepfakes is vital;

3. Develop AI Algorithms: Efforts should be made to reduce existing biases in AI algorithms that may skew content distribution towards mainstream content. Balancing the representation of both mainstream and niche content can foster a more diverse and inclusive film industry;

4. Strengthen the Education System: To enhance public understanding and discernment, educational programs should focus on cultivating critical thinking skills. This will empower individuals to critically evaluate the authenticity and implications of AI-generated content in filmmaking.

These recommendations aim to create a framework that promotes ethical use, transparency, diversity, and informed public engagement with AI technologies in the film industry.

### 5.3. Limitations and Future Research

Limitations:

1. Geographical Limitations: The current study focuses exclusively on English-language films, neglecting the rich and diverse AI practices in non-Western film industries such as Bollywood in India and Nollywood in Nigeria. This narrow geographical scope limits the generalizability of the findings to a global context;

2. Data Limitations: A significant constraint is the scarcity of available data, which prevents the study from providing concrete methods or metrics to quantify the extent of human versus AI involvement in film production processes. Without such quantitative measures, it becomes challenging to assess the precise impact of AI on filmmaking;

3. Ethical Complexity: The study does not address the ethical implications of audience acceptance of AI bylines. Specifically, there is no quantification of the threshold at which audiences might find AI credits acceptable or uncomfortable. This omission could have been addressed through an experimental psychology approach to gauge public sentiment and reactions more accurately.

Future Research:

1. Neurocinematology Experiment: Future studies should consider employing neurocinematology techniques, such as functional Magnetic Resonance Imaging (fMRI), to monitor amygdala activation levels in audiences when viewing AI-generated footage. This could provide insights into the "technical discomfort" associated with AI-produced content and help refine ethical guidelines for its use;

2. Optimization of Communication Algorithms: There is a need to explore and optimize communication algorithms, particularly those governing social media platforms. By rectifying these algorithms, independent filmmakers might be able to mitigate recommendation biases that currently favor mainstream content over niche or independent productions;

3. Human Control Index (HCI) Quantitative R&D: A critical area for future research is the development of a Human Control Index (HCI). This index should establish specific methods and metrics to quantify human and AI involvement in film production, enabling more precise assessments of their respective roles and contributions.

In refining these sections, I aim to enhance clarity, ensure that each point is well-articulated, and maintain a professional tone suitable for academic or research contexts. It's also important to highlight the significance of addressing these limitations and how the proposed future research can contribute to overcoming them.